\documentclass[]{interact}

\usepackage{epstopdf}
\usepackage{subfigure}

\usepackage{natbib}
\bibpunct[, ]{(}{)}{;}{a}{}{,}

\theoremstyle{plain}

\theoremstyle{definition}

\theoremstyle{remark}

\usepackage{color}
\usepackage{fancyvrb}

\DefineVerbatimEnvironment{Highlighting}{Verbatim}{commandchars=\\\{\}}
\usepackage{framed}
\definecolor{shadecolor}{RGB}{248,248,248}
\newenvironment{Shaded}{\begin{snugshade}}{\end{snugshade}}

\newcommand{\DataTypeTok}[1]{\textcolor[rgb]{0.13,0.29,0.53}{#1}}

\newcommand{\KeywordTok}[1]{\textcolor[rgb]{0.13,0.29,0.53}{\textbf{#1}}}
\newcommand{\NormalTok}[1]{#1}
\newcommand{\OperatorTok}[1]{\textcolor[rgb]{0.81,0.36,0.00}{\textbf{#1}}}

\usepackage{hyperref}
\usepackage[utf8]{inputenc}
\def\tightlist{}
\usepackage{setspace}

\begin{document}

\articletype{Article}

\title{Burning sage: Reversing the curse of dimensionality in the visualization
of high-dimensional data}

\author{\name{Ursula Laa$^{a, b}$, Dianne Cook$^{b}$, Stuart Lee$^{b, c}$}
\affil{$^{a}$School of Physics and Astronomy, Monash University; $^{b}$Department of Econometrics and Business Statistics, Monash University; $^{c}$Molecular Medicine Division, Walter and Eliza Hall Institute, Parkville,
Australia}
}

\thanks{CONTACT Ursula Laa. Email: \href{mailto:ursula.laa@monash.edu}{\nolinkurl{ursula.laa@monash.edu}}, Dianne Cook. Email: \href{mailto:dicook@monash.edu}{\nolinkurl{dicook@monash.edu}}, Stuart Lee. Email: \href{mailto:stuart.lee1@monash.edu}{\nolinkurl{stuart.lee1@monash.edu}}}

\maketitle

\begin{abstract}
In high-dimensional data analysis the curse of dimensionality reasons
that points tend to be far away from the center of the distribution and
on the edge of high-dimensional space. Contrary to this, is that
projected data tends to clump at the center. This gives a sense that any
structure near the center of the projection is obscured, whether this is
true or not. A transformation to reverse the curse, is defined in this
paper, which uses radial transformations on the projected data. It is
integrated seamlessly into the grand tour algorithm, and we have called
it a burning sage tour, to indicate that it reverses the curse. The work
is implemented into the tourr package in R. Several case studies are
included that show how the sage visualizations enhance exploratory
clustering and classification problems.
\end{abstract}


\doublespacing

\hypertarget{introduction}{%
\section{Introduction}\label{introduction}}

The term ``curse of dimensionality'' was originally introduced by
\citet{BellmanRichard1961}, to express the difficulty of doing
optimization in high dimensions because of the exponential growth in
space as dimension increases. A way to think about it is, that the
volume of the space grows exponentially with dimension, which makes it
infeasible to sample \emph{enough} points -- any sample will be less
densely covering the space as dimension increases. The effect is that
most points will be far from the sample mean, on the edge of the sample
space. \citet{doi:10.1111/j.1467-9868.2005.00510.x} have shown that in
the extreme case of high-dimension, low-sample size data, observations
are on the vertices of a simplex.

This affects many aspects of data analysis: minimizing the error during
model fitting relies on effective optimization techniques,
non-parametric modeling requires finding nearest neighbors which may be
far away and sampling from high-dimensional distributions is likely to
have points far from the population mean. \citet{Donoho00} considers the
curse of dimensionality as a blessing, because the sparsity can be
leveraged for computational efficiency. This is used in regularization
methods, like lasso, to penalize model complexity. The penalty term
results in shrinking (some of) the parameter estimates towards zero.

\begin{figure}

{\centering \includegraphics[width=0.95\linewidth]{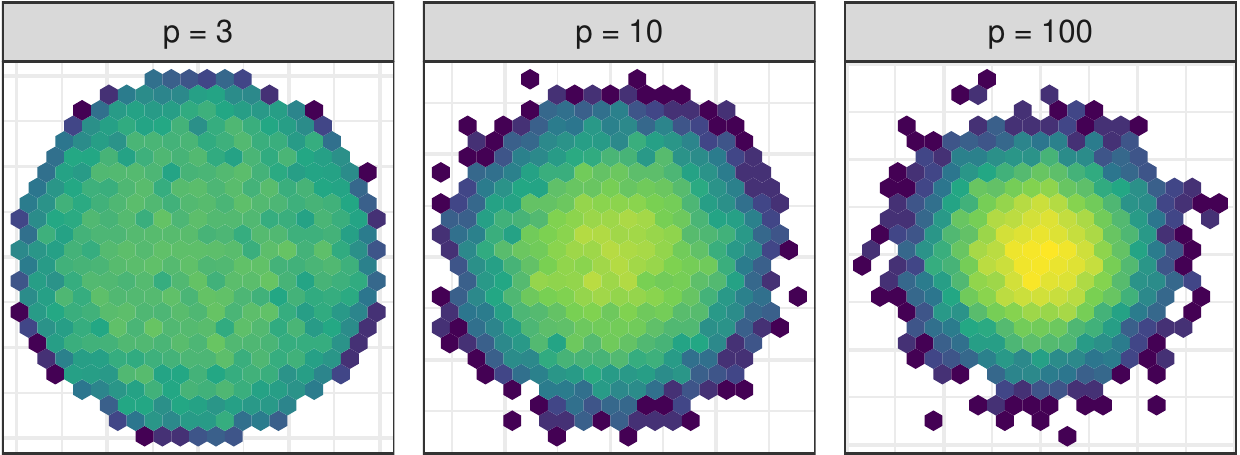} 

}

\caption{Illustration of data crowding, using hexbin plots of two-dimensional projections of 10k points sampled uniformly within $p$-dimensional hyperspheres, for $p=3, 10, 100$. The fill color shows the logarithm of the bin count, with the highest counts shown in yellow. As $p$ increases the density  concentrates near the center.}\label{fig:density}
\end{figure}

Paradoxically, the curse of dimensionality inverts for dimension
reduction, resulting in an excessive amount of observations near the
center of the distribution. This affects visualizations made on
low-dimensional projections, like the tour \citep[\citet{BCAH05}]{As85}.
The effect is described by \citet{diaconis1984}, that most
low-dimensional linear projections are approximately Gaussian, with
observations concentrating in the center. This has motivated the
development of indexes for projection pursuit which search for departure
from normality. It is also related to what is called ``data piling'' in
high-dimension low-sample size data
\citep[\citet{10.1093/biomet/asp084}]{10.2307/27639976}: all
observations can collapse into a single point. These issues also persist
with non-linear dimension reduction techniques, and are often referred
to as the ``crowding problem'', which methods like t-Distributed
Stochastic Neighbor Embedding (t-SNE) \citep{tsne} aim to alleviate.
Figure \ref{fig:density} illustrates the crowding problem.
Two-dimensional linear projections of points sampled uniformly within
\(p\)-dimensional hyperspheres (\(p=3, 10, 100\)) are displayed as
hexbin plots. Color indicates log count of the bin, with yellow mapping
the highest counts. As \(p\) increases the density concentrates in the
center of the projection.

In this work we address data crowding in low-dimensional linear
projections by providing a reverse transformation for tour methods.
Tours show interpolated sequences of low-dimensional projections of the
data. When exploring data with a tour we can discover features that are
only visible in linear combinations of variables. However, the data
crowding could hide these features. To reverse the effect, we introduce
a radial transformation that magnifies the center of the distribution.
This is called a burning sage tour, to reflect that the crowding caused
by the curse of dimensionality is being removed.

The paper is structured as follows. The radial transformation and its
implementation is described in Section \ref{sec:method}. Section
\ref{sec:application} illustrates the use of the sage tour with examples
in clustering, supervised classification and a classical
needle-in-the-haystack problem. Section \ref{sec:concl} describes
possible extensions to the method.

\hypertarget{sec:method}{%
\section{Burning sage algorithm}\label{sec:method}}

To understand why points tend to be away from the center in the
high-dimensional space, but crowd the center in low-dimensional
projections, it is helpful to consider the projected volume relative to
high-dimensional volume. To avoid edge effects and to impose rotation
invariance, we will start from the data being uniformly distributed in a
hypersphere, i.e.~all data points are within a specified distance from
the center. This makes calculations more tractable than assuming a
uniform distribution in hypercube (box).

\begin{figure}

{\centering \includegraphics[width=0.6\linewidth]{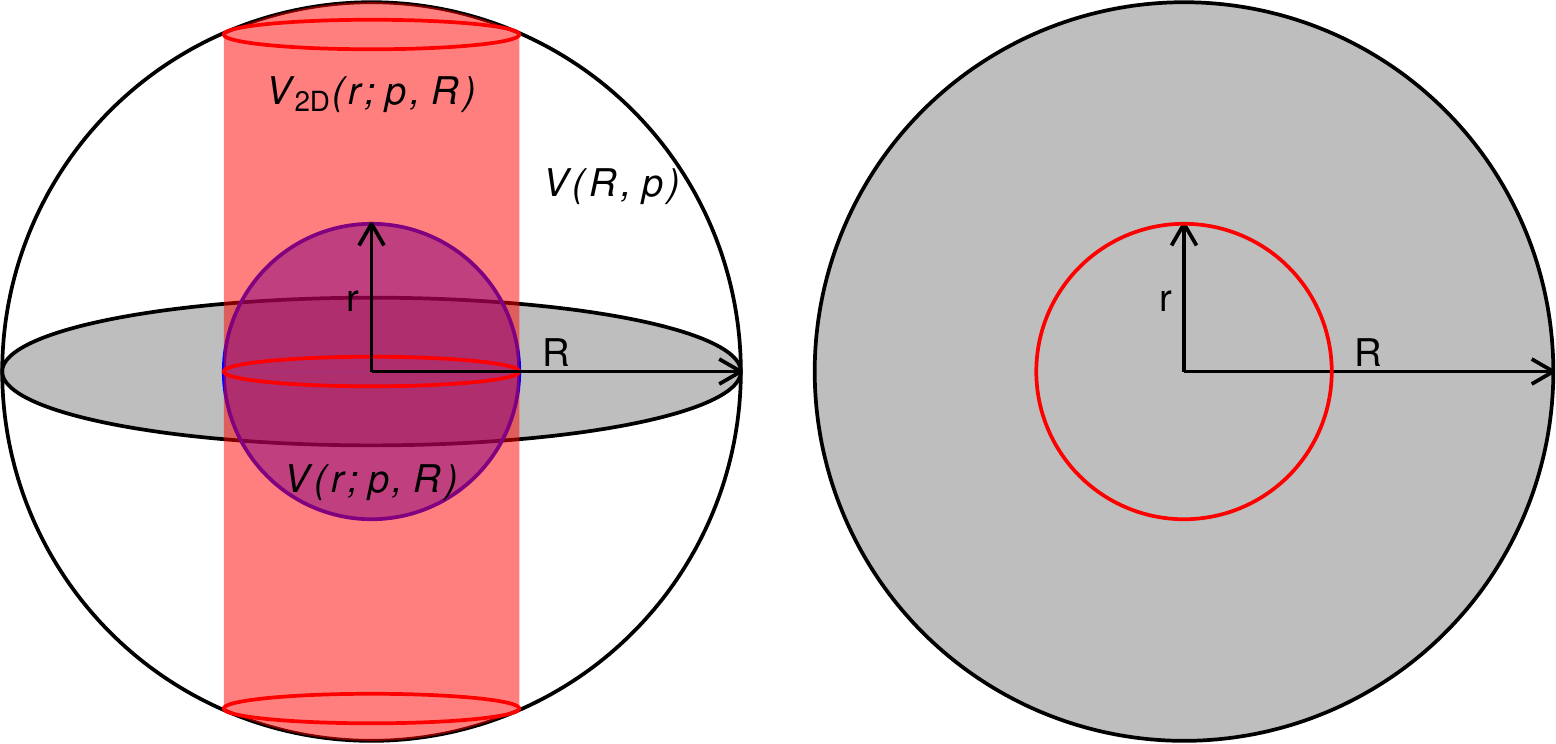} 

}

\caption{Illustration (and notation) for describing the elements used in the burning sage transformation. The 3D sphere (left) shows the different volumes to be compared. The full sphere has volume $V(R, p)$. Within a radius $r$ the sphere contains the reduced volume $V(r; p, R)$, shown in blue, but the projected volume within a radius $r$ in a two-dimensional plane is much larger, given by the volume of the cylinder with rounded caps, $V_{2D}(r; p ,R)$, shown in red. The intersection of the plane with the sphere is illustrated in grey, and the plane representing the projection with both radii is shown at right.}\label{fig:sketch}
\end{figure}

Figure \ref{fig:sketch} illustrates the comparison to be made, using
something we can easily picture, a 3D sphere. Projecting the data from
within a 3D sphere to 2D (grey disk) will result in mass being condensed
into the disk. Imagine comparing the volume of a cylinder at different
locations in the disk. A centered cylinder has more volume. This is
exaggerated as \(p\) increases: the centered cylinder has much more
volume than any other cylinder.

To reverse this effect, we introduce a radial transformation that
redistributes the projected points, such that equal volume in the
original (\(p\)-dimensional) space is projected onto equal area in a
2-dimensional projection. Note that this can be generalized for
\(d\)-dimensional projections by mapping onto equal \(d\)-dimensional
volume instead.

\hypertarget{definition-of-the-relative-projected-volume}{%
\subsection{Definition of the relative projected
volume}\label{definition-of-the-relative-projected-volume}}

To understand how the \(p\)-dimensional volume is projected onto a
\(2\)-dimensional plane, we study what fraction of the total volume is
projected onto the area of a disk depending on its radius. This
dependence was described in \citet{Laa:2020wkm}. We start from a
\(p\)-dimensional hypersphere, with radius \(R\) and volume \(V(R, p)\),
and its projected volume onto a centered \(2\)-dimensional disk of
radius \(r\), \(V_{2D}(r; p, R)\), where \(r\) can be any radius within
\([0, R]\). The relative projected volume is then given as the ratio of
these two quantities, \begin{equation}
v_{2} (r; p, R) = \frac{V_{2D}(r; p, R)}{V(R, p)} = 1 - \left(1-\left(\frac{r}{R}\right)^2\right)^{p/2}.
\label{eq:cdf}
\end{equation} This ratio is of particular interest because it gives the
2-dimensional radial cumulative distribution function (CDF) of points
when assuming a uniform distribution within the \(p\)-dimensional
hypersphere.

We can compare \(v_{2} (r; p, R)\) to the relative volume within a
radius \(r\) in the original \(p\)-dimensional hypersphere,
\begin{equation}
v_{p} (r; p, R) = \frac{V(r, p)}{V(R, p)} = \left({\frac{r}{R}}\right)^p.
\label{eq:vp}
\end{equation}

\begin{figure}

{\centering \includegraphics[width=0.9\linewidth]{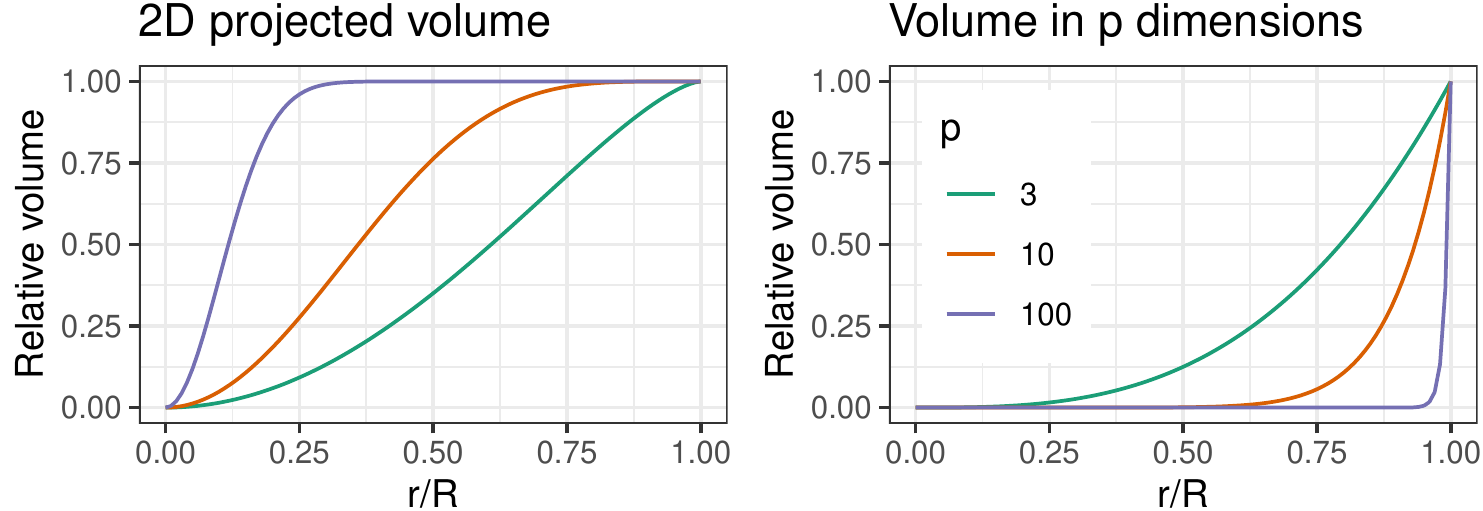} 

}

\caption{Comparing relative volume of a $p$-dimensional hypersphere captured within a radius $r$, in the 2-dimensional projection (left) and in the $p$-dimensional space (right), for $p=3, 10, 100$. The difference is dramatic, which illustrates the paradox of the curse of dimensionality. The relative volume of a $p$D sphere shrinks, as $p$ increases, while the projected volume (near the center) grows.}\label{fig:cdf}
\end{figure}

Figure \ref{fig:cdf} compares these two quantities (Eq. \ref{eq:cdf} and
\ref{eq:vp}), for \(p=3, 10, 100\). On the left is \(v_{2} (r; p, R)\)
and on the right is \(v_{p} (r; p, R)\). The function shapes change in
opposite directions as \(p\) increases: \(v_{2} (r; p, R)\) peaks
earlier and \(v_{p} (r; p, R)\) gets flatter. This is the paradox of the
curse of dimensionality in that projected volume at the center increases
with \(p\).

\hypertarget{calculating-the-radial-transformation}{%
\subsection{Calculating the radial
transformation}\label{calculating-the-radial-transformation}}

The aim of the algorithm is to redistribute the projected volume such
that equal relative areas on the disk, as given by
\(v_{2} (r; p=2, R)= v_p(r; p=2, R) = (r/R)^2\), contain equal relative
projected volume, given by \(v_{2} (r; p, R)\). This is achieved through
a transformation of the projected radius that can be defined for any
\(r\in[0,R]\), and is applied to the projected data points in the plane,
\(y = (y_1, y_2)\). We work with polar coordinates and represent the
data points as \(y = (r_y, \theta_y)\). The angular component
\(\theta_y\) is uniform for this distribution, by the rotation
invariance of the sphere, and thus does not need to be transformed. The
radial component \(r_y\) is transformed in two steps.

The first transformation is to replace \(r_y\) with
\(v_{2} (r_y; p, R)\). Since this is the radial CDF of the assumed
underlying distribution, we expect that \(v_{2} (r_y; p, R)\) is
approximately uniformly distributed in radius. We then transform
\(v_{2} (r_y; p, R)\) using the inverse of \(v_{2} (r_y; 2, R)\), to go
from a uniform distribution in radius to a uniform distribution in area
of the disk. This inverse is defined via \begin{equation}
v_2^{-1}(v_2(r_y; 2, R); 2, R) = v_2(v_2^{-1}(r_y; 2, R); 2, R) = r_y
\end{equation} and thus \begin{equation}
v_2^{-1}(r_y; 2, R) = R \sqrt{r_y}.
\end{equation}

The full radial transformation is therefore given by \begin{equation}
r'_y = v_2^{-1} (v_2(r_y; p, R); 2, R) =  R \sqrt{v_2(r_y; p, R)} = R \sqrt{1-\left(1-\left(\frac{r_y}{R}\right)^2\right)^{p/2}}
\label{eq:resc}
\end{equation}

\begin{figure}

{\centering \includegraphics[width=0.6\linewidth]{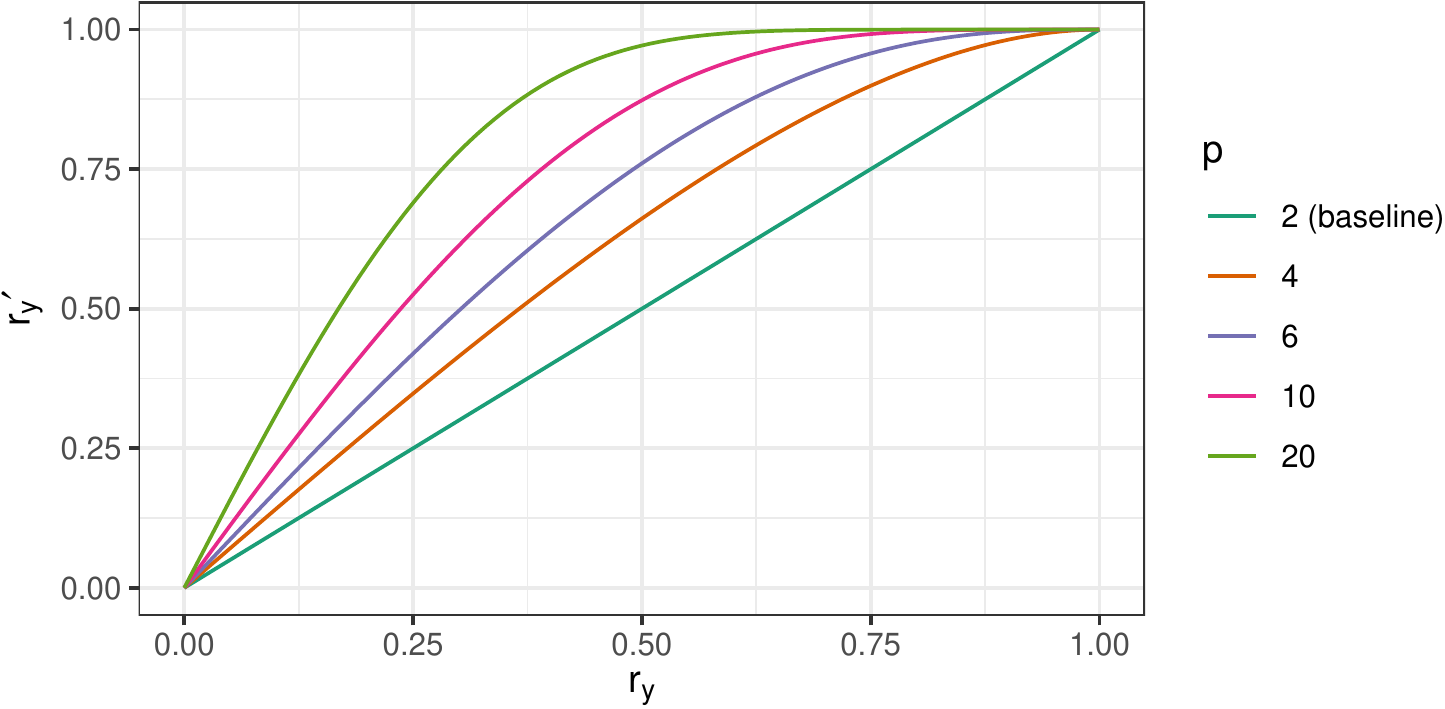} 

}

\caption{Relation between $r_y$ and $r'_y$ for different values of $p$ and assuming $R=1$. The scaling is approximately linear near the center, but leads to distortion at large radii when $p$ is large.}\label{fig:radii}
\end{figure}

\begin{figure}

{\centering \includegraphics[width=1\linewidth]{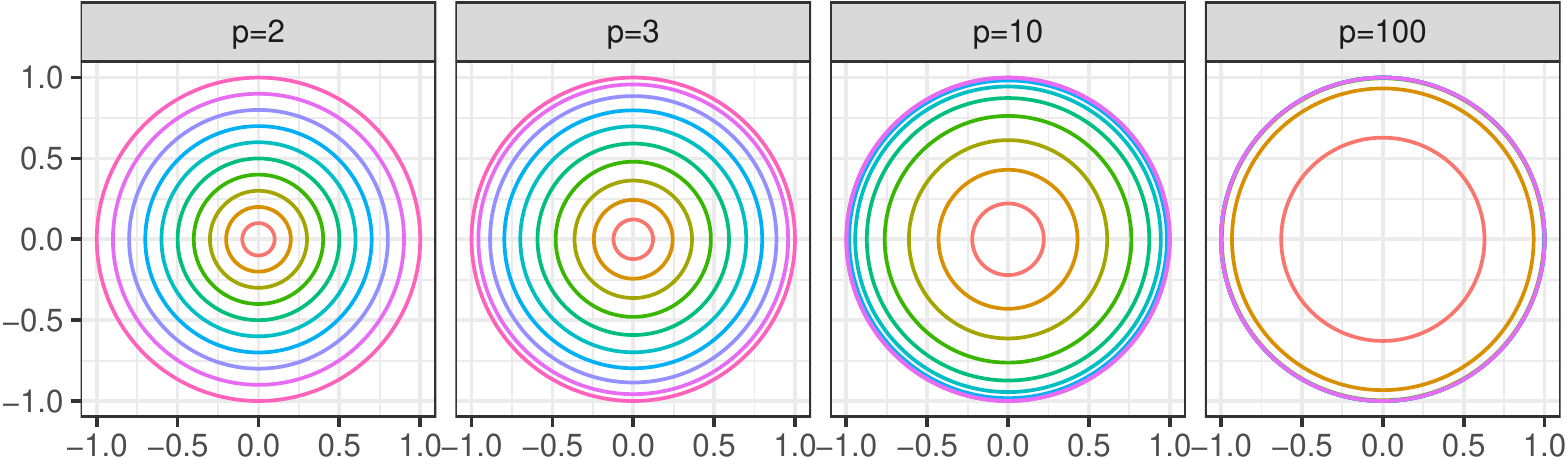} 

}

\caption{Equidistant concentric circles, for $p=2, 3, 10, 100$, illustrating the radial transformation. The circles remain equidistant for $p=2$ where no transformation is performed, and get pushed out towards the edge as $p$ increases.}\label{fig:circles}
\end{figure}

The relation between \(r'_y\) and \(r_y\) depends on the number of
dimensions \(p\), and is illustrated for selected values in Figure
\ref{fig:radii}. We see that the transformation is approximately linear
near the center. As \(p\) increases it becomes non-linear faster, and
for \(p=10\), for example, the points with radius \(r_y>0.5\) will
already be highly distorted and pushed out towards the last eighth in
\(r'_y\). Figure \ref{fig:circles} demonstrates this for different
values of \(p\) by showing equidistant circles for which the radius has
been transformed according to Eq. \ref{eq:resc}.

\hypertarget{sec:params}{%
\subsection{Trimming and tuning}\label{sec:params}}

The transformation in Eq. \ref{eq:resc} is fixed for a given input
dataset, by evaluating the number of dimensions \(p\) and the maximum
distance from the center \(R\). However, in practice we may wish to trim
the projected data or tune the transformation. A combination of both
adjustments can be used to further zoom in on the center of the
distribution, or alternatively, to soften the transformation.

\hypertarget{trimming}{%
\subsubsection{Trimming}\label{trimming}}

The overall scale of the transformation is determined by \(R\). In the
case of an approximately spherical and uniform distribution the maximum
distance from the center works well and ensures the validity of the
rescaling in Eq. \ref{eq:resc}. But this is not robust and might result
in a much larger scale than desired, especially when it is determined by
outlying observations.

We therefore allow trimming of the projected observations, using \(R\)
as a free parameter of the display function. When selecting a value
\(R\) that is smaller than the maximum distance from the center, we need
to ensure that the projected radius of points is always smaller than
\(R\), by trimming \(r_y\) as \begin{equation}
r_y^{\mathrm{trim}} = \min(r_y, R)
\label{eq:cutR}
\end{equation} for each observation.

\hypertarget{tuning}{%
\subsubsection{Tuning}\label{tuning}}

The dimension of the input might not reflect the intrinsic
dimensionality of the dataset. This could be the case when dimension
reduction was used prior to visualization, e.g.~displaying only the
first few principal components. In this case the effective
dimensionality \(p_{\mathrm{eff}}\) is likely between the original
number of dimensions and the selected number of principal components. We
can think of omitted components as being in the orthogonal space of all
considered projections, with some directions being pure noise, while
others may still carry relevant information.

We allow tuning \(p_{\mathrm{eff}} = \gamma p\) by selecting the scaling
parameter \(\gamma\). By default, \(\gamma=1\) and
\(p_{\mathrm{eff}}=p\). When \(\gamma<1\) the rescaling will be softer,
and \(\gamma>1\) results in more aggressive rescaling than suggested by
\(p\) alone. Note that when \(p_{\mathrm{eff}} < 2\) we actually invert
the behavior and shift the focus away from the center, in general this
is not recommended.

\hypertarget{sec:implementation}{%
\subsection{Implementation as a dynamic
display}\label{sec:implementation}}

While the radial transformation can in general be used with any
low-dimensional display that suffers from data crowding, it is most
useful when combined with a dynamic display showing a sequence of
interpolated low-dimensional projections obtained when running a tour.
We have implemented it as a new display method called
\texttt{display\_sage} in the \texttt{tourr} package \citep{tourr} in R
\citep{rref}.

We can think of the display functions as part of a data pipeline
obtained when running a tour. The initial step is pre-processing the
data, given by \(\mathbf{X}\), an \(n \times p\) matrix containing \(n\)
observations in \(p\) dimensions. Typically, this includes centering and
scaling, using either the overall range or the variance. Ensuring a
common scale of all variables, comparable to the selected scale
parameter \(R\), is especially important with the new display. The tour
then iterates over the following steps:

\begin{enumerate}
\def\labelenumi{\arabic{enumi}.}
\tightlist
\item
  Obtain projection matrix \(\mathbf{A}\). For \(d\)-dimensional
  projections this is an orthonormal \(p \times d\) matrix. To ensure
  the smooth rotation of projections, each new \(\mathbf{A}\) is
  obtained as an interpolated step in the sequence, as explained in
  \citet{BCAH05}.
\item
  Project the data by computing
  \(\mathbf{Y} = \mathbf{X}\cdot\mathbf{A}\).
\item
  Map \(\mathbf{Y}\) to the display to re-draw the projected data. For
  \(d=2\) this typically maps the projected points onto a scatter plot
  display. With the new display we first transform \(\mathbf{Y}\) as:

  \begin{itemize}
  \tightlist
  \item
    Center the 2-dimensional matrix \(\mathbf{Y}\) and compute its polar
    coordinate representation \((r_y, \theta_y)\).
  \item
    For each observation, first use Eq. \ref{eq:cutR} to get the trimmed
    radius \(r_y^{\mathrm{trim}}\) within the specified range, and then
    apply the radial transformation defined in Eq. \ref{eq:resc} to
    obtain \(r'_y\).
  \item
    Use the transformed radial coordinate \(r'_y\) to re-compute the
    mapping onto Euclidean coordinates \((y_1, y_2)\).
  \end{itemize}
\item
  To fit the final projection onto the plotting canvas ranging between
  \([-1,1]\), we rescale each mapped observation \(y\) using a scaling
  parameter \(s\).
\end{enumerate}

The display can be added when calling the \texttt{animate} function in
\texttt{tourr}, as

\begin{Shaded}
\begin{Highlighting}[]
\NormalTok{tourr}\OperatorTok{::}\KeywordTok{animate}\NormalTok{(}
\NormalTok{  data,}
  \DataTypeTok{tour_path =}\NormalTok{ tourr}\OperatorTok{::}\KeywordTok{grand_tour}\NormalTok{(),}
  \DataTypeTok{display =} \KeywordTok{display_sage}\NormalTok{(gam, R, half_range)}
\NormalTok{  )}
\end{Highlighting}
\end{Shaded}

\noindent and uses \texttt{gam} to set the \(\gamma\) parameter for
computing \(p_{\mathrm{eff}}\), and the overall range \texttt{R} used
for trimming. Both these parameters are described in Section
\ref{sec:params} above. Finally, \texttt{half\_range} sets the scale
parameter \(s\) to adjust the scale to the drawing canvas. The ratio
\(R/s\) sets the scale for fitting the displayed data on the plotting
canvas, by default \(s = R\) and we apply the scaling factor \(0.9\) to
contain the projected points within the display. When adjusting \(R\)
the user should take care to adjust \(s\) accordingly.

\hypertarget{sec:application}{%
\section{Applications}\label{sec:application}}

To illustrate the benefit of using the reverse transformation for
examining data using a tour, four applications are shown: clustering of
single cell RNA-seq, classifying hand-sketched images, comparing physics
experiments, and the classical pollen data. The pollen data example is
used to illustrate the effect of parameter choices in the sage tour.

\hypertarget{sec:appl1}{%
\subsection{Clustering single-cell RNA-seq data sets}\label{sec:appl1}}

In the analysis of single cell RNA-seq data, cluster analysis is
performed to detect cell types and characterize the expression of genes
that define those cell types, and the relative orientation of the cell
types to each other (trajectory analysis) \citep{Amezquita2020-at}.
Generally, for cluster verification, analysts use embedding methods like
t-SNE to verify the placement and meaning of clusters from a clustering
algorithm. An alternative is to use a tour on a small number of
principal components to examine the clusters relative to gene
expression.

Here we compare the sage display and regular tour display on mouse
retinal single cell RNA-seq data from \citet{Macosko2015-ot}. The raw
data consists of a 49,300 cells and was downloaded using the using the
\texttt{scRNAseq} Bioconductor package \citep{scRNAseq-d}. We use a
standard workflow for pre-processing and normalizing this data
(described by \citet{Amezquita2020-at}): quality control was performed
using the \texttt{scater} package \citep{McCarthy2017} and
\texttt{scran} \citep{Lun2016} was used to transform and normalize the
expression values and select highly variable genes (HVGs). The top ten
percent of the most HVGs were used as features to subset the normalized
expression matrix and compute the principal components. Using the first
25 PCs we built a shared nearest neighbors graph (with \(k = 10\)) and
clustered this graph using Louvain clustering, resulting in 11 clusters
being formed \citep{Blondel2008-bx}.

\begin{figure}

{\centering \includegraphics[width=0.9\textwidth]{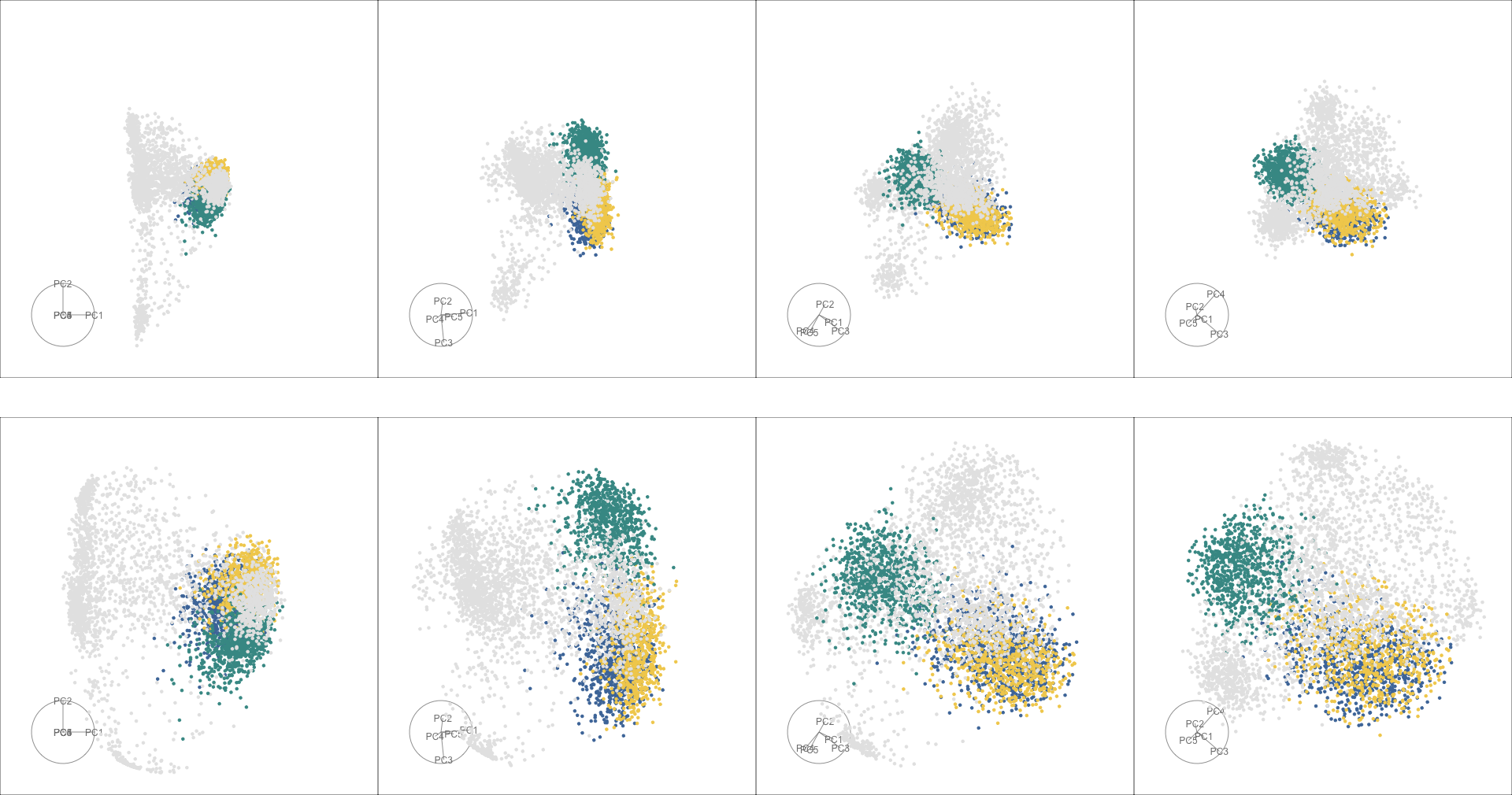} 

}

\caption{Selected frames from using a tour of the mouse data with the default tour display (top), and the sage display with $\gamma=3$ (bottom). Three selected clusters are highlighted in color, all other points are shown in grey. Using the sage display mitigates overplotting and provides a better understanding of cluster separation.}\label{fig:mouse-cluster}
\end{figure}

A tour is run on the first five PCs (approximately 20\% of the variance
in expression), on a weighted subsample of cells based on their cluster
membership - 4,590 cells. For the sage display, we set \(\gamma = 3\),
fixing the effective dimensionality of the data to
\(p_{\mathrm{eff}} = 15\). The PCs are scaled to have zero mean and unit
variance. Here we focus on comparing three of the clusters. In the PC
plots they look very similar, begging the question whether they should
be considered to be separate groups. Figure \ref{fig:mouse-cluster}
shows selected frames from a default tour (top row) and the sage tour
(bottom row). The columns show the same projection, with the difference
being that the sage transformation is applied in the sage tour
projections. The full animations are available in the supplementary
material. The static plots serve to illustrate the main points, but we
encourage the reader to look at the tour animations to fully appreciate
the advantage of the sage display.

Using the default tour display (Figure \ref{fig:mouse-cluster}, top),
the three clusters (dark green, blue, and yellow) are obscured by points
in other clusters as we move through the frames of the animation. The
points in the dark green cluster are overlapping those found in the
yellow and blue clusters; and it is difficult to see if there is any
separation between the blue and yellow clusters. In contrast, the sage
display (Figure \ref{fig:mouse-cluster}, bottom), expands the center of
projection, and results in the differences between the three clusters
being more visible. Particularly, the relative positions of the yellow
and blue clusters are easier to see. While these clusters are distinct
from the dark green cluster, in most frames they are still overlapping
and mixed together, providing evidence that it may be appropriate to
consider them a single cluster. Conversely, it can be seen that the dark
green cluster is distinctly separated from the other two in some
projections. The sage tour makes these comparisons a little easier.

\hypertarget{sec:appl2}{%
\subsection{Classifying hand-sketches}\label{sec:appl2}}

We next use the new display to look at different distributions of images
from the Google QuickDraw collection \citep{quickdraw-paper}. These are
\(28\times28=784\) pixel grey scale data that are available publicly. In
this example, we sample 1000 images from three types of sketches
(banana, cactus, crab) and see if we can separate the classes in the
high-dimensional parameter space.

We reduce the dimensionality from 784 variables to the first 5 PCs,
which captures approximately 20 per cent of the variation of the data.
Before applying the tour we rescale each component to have mean zero and
unit variance. To account for the dimension reduction before
visualization we set \(\gamma=2\) for the sage display.

\begin{figure}

{\centering \includegraphics[width=0.9\textwidth]{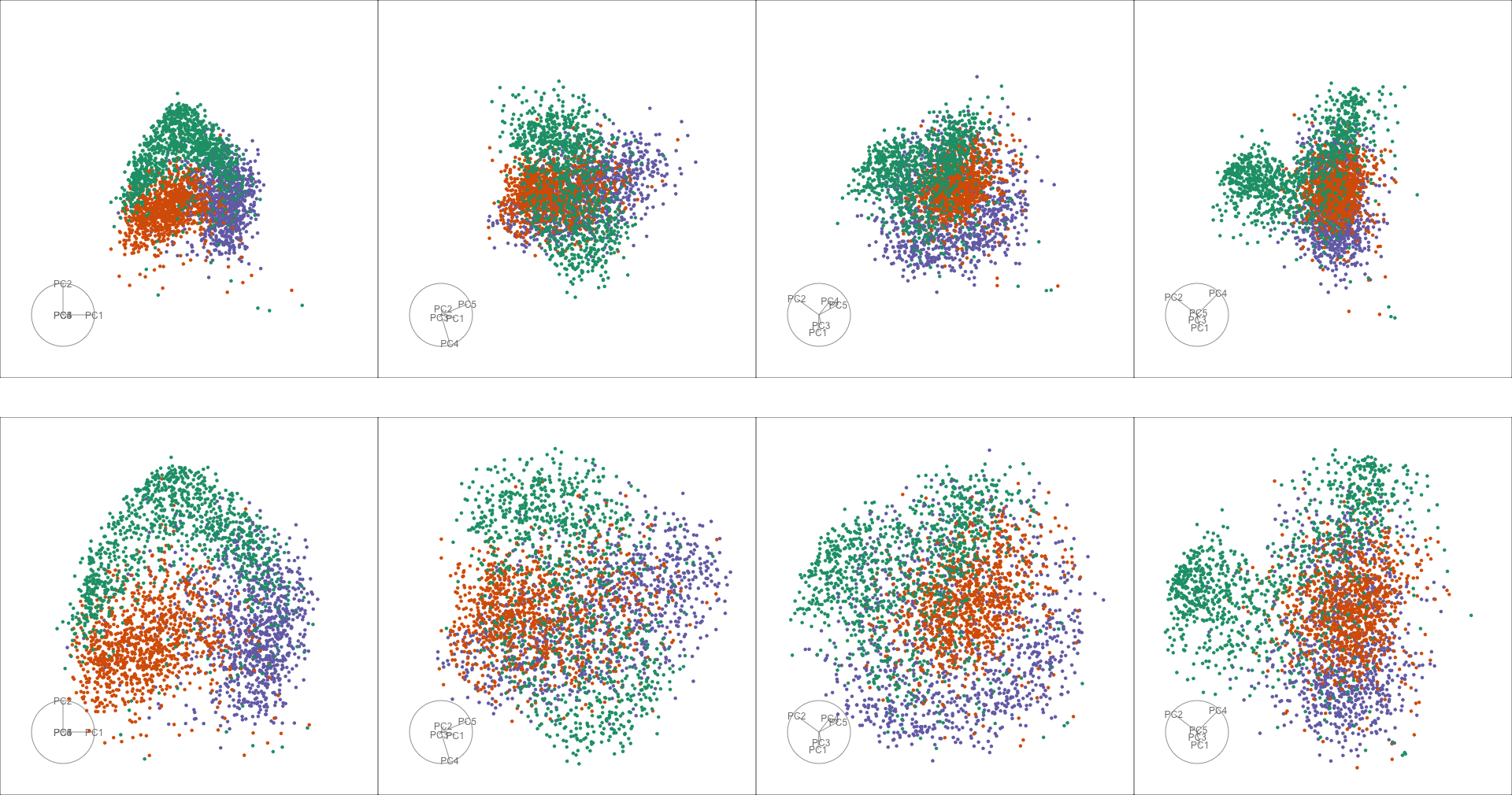} 

}

\caption{Selected frames  of the tour run on the sketches data using the default tour display (top), and using the sage display with $\gamma=2$ (bottom). Three types of sketches are indicated by color: banana (green), cactus (orange) and crab (purple). Overplotting of points is a problem for the grand tour display, while the sage display reveals low density near the center.}\label{fig:sketches}
\end{figure}

Figure \ref{fig:sketches} shows the grand tour on the PCs, where green
points correspond to the banana class, orange points represent the
cactus class and purple points are the crab class. In the selected
frames of both displays points belonging to the cactus class are
concentrated near the center, however on the default display (Figure
\ref{fig:sketches}, top) there is overplotting: points from other
classes overlap those in the cactus class. The sage display (Figure
\ref{fig:sketches}, bottom) helps reduce overplotting, it is easier to
see that the centers of class are separated and that there is
substructure in the banana class, which further collapses into two
subgroups.

The animated sage tour available in the supplementary material further
reveals a low density of points near the center of the distribution:
observing the movement of points when rotating the viewing angle shows
that even the cactus class is clustering away from the mean.

\hypertarget{sec:appl3}{%
\subsection{Comparing physics experiments: PDFSense}\label{sec:appl3}}

Data were obtained from CT14HERA2 parton distribution function fits and
describe the sensitivity of fit parameters to experimental measurements
\citep{Wang:2018heo}. There are 28 parameters, and varying one at a time
to move \(\pm 1 \sigma\) away from the `best fit point' (maximum
likelihood estimate) provides our input variables, labelled X1-X56. Each
of the 2808 observations corresponds to a physical observable and
measures how the fit prediction changes along the 56 directions in
parameter space. Points are grouped based on the underlying process in
the experiment, which is mapped to color in the following. With the
analysis of the distribution along these variables X1-X56 we can
understand to what extent each experimental measurement provides new
information for the global fit. For example, orthogonality between
groups marks complementary constraints, and outlying points are
considered as important for future fits, see discussion in
\citet{Cook:2018mvr}.

\begin{figure}

{\centering \includegraphics[width=0.9\textwidth]{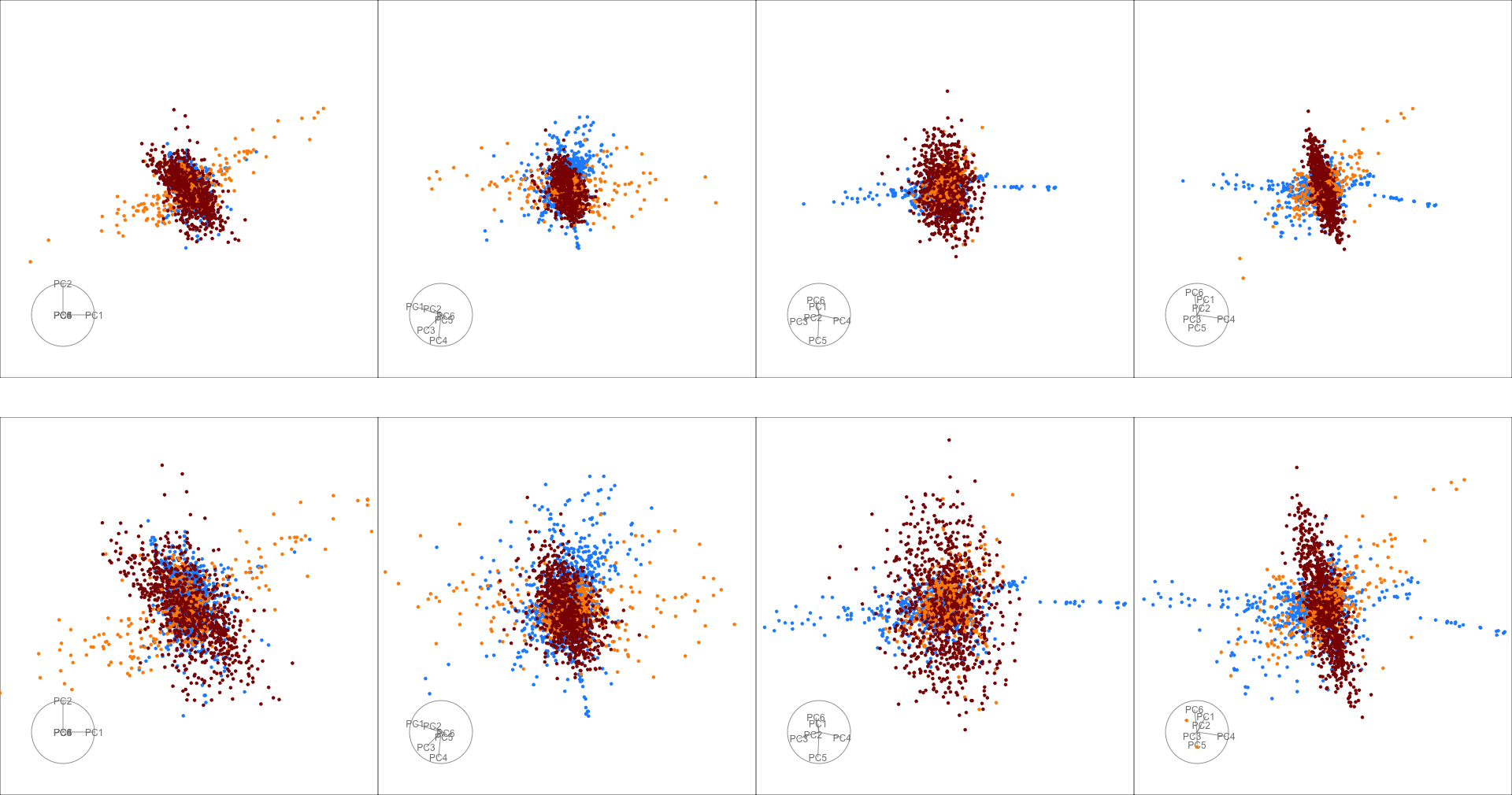} 

}

\caption{Selected frames of the tour of the pdfsense data using the default tour display (top), and using the sage display with $R=10$ (bottom). Different underlying physical processes are shown by color and we can see orthogonality between the three groups. The sage display preserves the overall structure while revealing details that are hidden near the center in the default display.}\label{fig:pdfsense}
\end{figure}

Following the processing described there, we tour the first 6 PCs,
rescaled to have zero mean and unit variance. In Figure
\ref{fig:pdfsense} we see that the sage display with \(R = 10\) (Figure
\ref{fig:pdfsense}, bottom), maintains the overall shape of the data
seen using the default tour display (Figure \ref{fig:pdfsense}, top).
The different physical process, shown in different colors, are indeed
orthogonal in the parameter space, as can be seen most clearly by
looking at the animations available in the supplementary material.

The particular structure of this distribution, with some clusters
extending linearly away from the center and a set of outlying points,
results in poor use of the plotting space, and high level of clustering
near the center. For example, focusing on the blue cluster, we can see
that it extends out along different directions, but it can be
challenging to observe how the points move under the tour rotation, as
overplotting becomes an issue when points move through the center. Here,
the new display (bottom row) shows a clearer view.

\hypertarget{sec:appl4}{%
\subsection{Tuning the parameters: Pollen}\label{sec:appl4}}

The classical pollen data is useful to demonstrate the trimming and
tuning parameters. The five-dimensional data set was simulated by David
Coleman of RCA Labs, for the Joint Statistics Meetings 1986 Data Expo
\citep{pollen}, and is an example of a hidden structure near the center
of a distribution. The data are standardized by centering and scaling
such that the standard deviation of each variable is equal to one.

\begin{figure}
\centering
\includegraphics[width=0.3\textwidth]{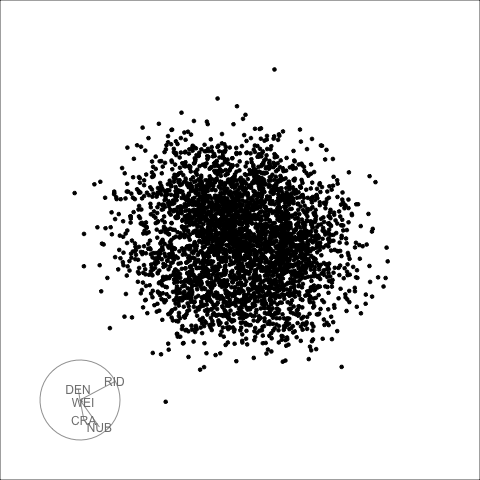}
\includegraphics[width=0.3\textwidth]{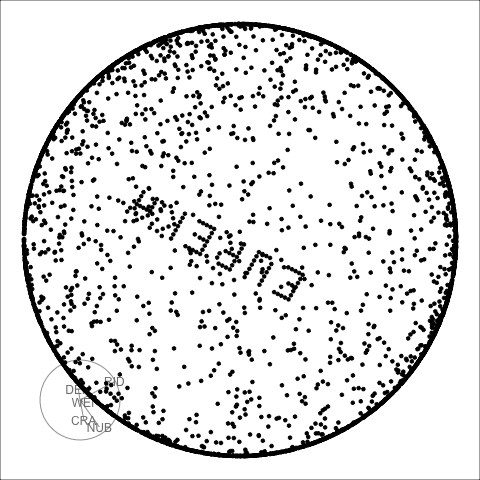}
\includegraphics[width=0.3\textwidth]{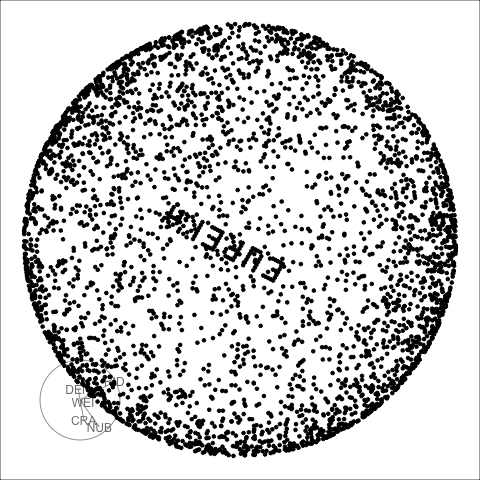}
\caption{Selected views of the pollen data in the new sage display, with default settings (left), setting $R=1$ (middle) and $\gamma=20$ (right). We can tune either $\gamma$, $R$ or a combination of the two to reveal the word "EUREKA" near the center of the distribution.}
\label{fig:pollen}
\end{figure}

Neither the standard tour display nor the sage display with default
settings (\(R=6.6\) which is set by the data scale, and \(\gamma=1\))
reveals the structure (left plot in Figure \ref{fig:pollen}). We can use
either \(\gamma\), \(R\) or a combination of the two to zoom in further
near the center. For example, we can use trimming \((R=1, \gamma=1)\)
(middle plot) or tuning \((R=6.6,\gamma=20)\) (right plot) as shown in
Figure \ref{fig:pollen}. There is an approximate equivalence between the
results obtained using either tuning or trimming, and both views clearly
reveal the word ``EUREKA'' hidden in the distribution.

While the static views look very similar, comparing the tour animations
(available in the supplementary material) reveals some differences
between the display with trimming or tuning. When trimming (by setting
\(R=1\)) the focus is clearly on the center of the distribution, and
most points get pushed out towards a maximum radius circle. On the other
hand, tuning the display by setting \(\gamma=20\) preserves the
elliptical shape of the distribution, making it easier to see
correlation patterns.

\hypertarget{sec:concl}{%
\section{Discussion}\label{sec:concl}}

This paper has introduced the sage tour, which reduces the data crowding
effects that occur when taking low-dimensional projections of
high-dimensional data. This new technique is easily incorporated into
exploratory high-dimensional data analysis, and applications shown in
Section \ref{sec:application} provide examples of the following tasks:

\begin{itemize}
\tightlist
\item
  clustering: the sage display uncovered clusters that were originally
  obscured by data piling, while still giving the viewer an accurate
  assessment of the size of a cluster, and their relative orientation,
  as shown in the single cell RNA-seq example (Section \ref{sec:appl1})
\item
  classification: the sage display decreases the number of overlapping
  points between classes and provides better visual separation between
  classes compared to the regular tour, as shown in the sketches example
  (Section \ref{sec:appl2})
\item
  shape analysis: the sage display helps us understand structures across
  multiple dimensions, for example orthogonality between multiple
  groups, as shown in the pdfsense example (Section \ref{sec:appl3})
\item
  needle discovery: the sage display allows to find hidden signal that
  is concealed by the density of points around the center of the
  projection, as shown in the pollen example (Section \ref{sec:appl4})
\end{itemize}

The approach provides interpretable visualization that captures high
dimensional information and preserves global structure, and it is
complementary to non-linear dimension reduction techniques. For example,
when visualizing clusters, the sage display enables an assessment of
cluster shapes, and accurately captures relative position and
orientation. The burning sage transformation is global and does not
magnify local structure like t-SNE does.

An alternative is the slice tour \citep{sliceTour} which allows
distributions of points around the center of the data to be explored
using sections instead of projections. The slice tour is useful when
there are large numbers of observations or if there is concave structure
in the data.

The tuning parameters can be used to more aggressively expand the center
of the display. All of the examples shown had some tuning. The last
example demonstrated how points away from the projected center get moved
to the edge of the hypersphere as \(\gamma\) is increased or \(R\) is
decreased. With more center magnification, the non-linear transformation
can introduce distortions, but this is a well known problem for any
non-linear dimension reduction technique including t-SNE. However,
unlike t-SNE, any distortion introduced by the sage display is
interpretable because it is controlled by a simple function (Eq.
\ref{eq:resc}).

The sage display is fast to compute, which lends itself to being
embedded into an interactive interface. An ideal interface would allow
real time changes to the parameters of the transformation. This would be
especially useful when coupled with linked brushing in complementary
views.

\hypertarget{acknowledgements}{%
\section*{Acknowledgements}\label{acknowledgements}}
\addcontentsline{toc}{section}{Acknowledgements}

The authors gratefully acknowledge the support of the Australian
Research Council. The paper was written in \texttt{rmarkdown}
\citep{rmarkdown} using \texttt{knitr} \citep{knitr}.

\hypertarget{supplementary-material}{%
\section*{Supplementary material}\label{supplementary-material}}
\addcontentsline{toc}{section}{Supplementary material}

The source material for this paper is available at
\url{https://github.com/uschiLaa/burning-sage}. The animated gifs for
all applications are also included in html files in the supplementary
material.

\bibliographystyle{tfcad}
\bibliography{biblio.bib}

\end{document}